\documentclass[aps,prl,twocolumn,showpacs,groupedaddress]{revtex4-1}
\usepackage{graphicx}
\usepackage{epstopdf}
\usepackage{color}
\usepackage{url}
\usepackage{hyperref}
\usepackage[T1]{fontenc}
\usepackage[latin1]{inputenc}
\usepackage{float}
\usepackage{ulem}

\newcommand{\Ca}{$^{40}\mathrm{Ca}^+$~}
\newcommand{\nounderline}[1]{#1}

\def\figref{Fig.~}
\def\eqnref{Eq.~}
\let \mr=\mathrm

\usepackage{amsmath}	
\begin{document}
\title{Cavity-induced anti-correlated photon emission rates of a single ion}

\author{Hiroki Takahashi}
\author{Ezra Kassa}
\email{Ezra.Kassa@sussex.ac.uk}
\author{Costas Christoforou}
\author{Matthias Keller}
\affiliation{Department of Physics and Astronomy, University of Sussex, Brighton, BN1 9RH, United Kingdom}

\begin{abstract}
  
We report on the alteration of photon emission properties of a single trapped ion coupled to a high finesse optical fiber cavity. We show that the vacuum field of the cavity can simultaneously affect the emissions in both the infrared (IR) and ultraviolet (UV) branches of the $\Lambda-$type level system of \Ca despite the cavity coupling only to the IR transition. The cavity induces strong emission in the IR transition through the Purcell effect resulting in a simultaneous suppression of the UV fluorescence.
The measured suppression of this fluorescence is as large as 66\% compared with the case without the cavity.  
Through analysis of the measurement results, we have obtained an ion-cavity coupling of $\bar{g}_0 = 2\pi\cdot (5.3 \pm 0.1)$ MHz, the largest ever reported so far for a single ion in the IR domain.

\end{abstract}

\maketitle

The effect of a structured environment on the spontaneous emission of atomic particles was first discovered by Purcell \nounderline{in 1946 \cite{Purcell:46} and is named after him.} \nounderline{More than 30 years later}, {it was experimentally demonstrated} with Rydberg atoms \cite{goy1983observation,Hulet:85}.
Since then, enhanced/reduced emission rates due to an optical cavity surrounding single emitters have been shown in a variety of physical systems such as trapped ions \cite{Keller:04, kreuter2004spontaneous, Steiner:13}, semiconductor quantum dots \cite{Unsleber:15}, nitrogen-vacancy centers in diamond \cite{Wolf:15} and rare-earth ions in solids \cite{Karaveli:10, Ding:16}.
In particular, single trapped ions coupled to optical cavities provide an ideal environment to study and exploit the enhanced light-matter interaction \cite{Keller:04,stute2012tunable} due to their unparalleled quantum control. 
In these systems the use of cavities with small mode volumes is crucial in the enhancement of the ion-cavity coupling, and consequently the emission into the cavity mode. To this end, miniaturized fiber-based Fabry-P\'erot cavities (FFPCs) have been introduced and successfully combined with ion traps \cite{Steiner:13,Ballance:16}. 
The integration of ion traps and FFPCs recently allowed the Purcell effect to be studied extensively in a two-level system  \cite{Ballance:16}. However, the Purcell effect in a multi-level atomic system and its role on competitive transitions rates has not been investigated so far. In this letter we demonstrate the coupling of a single ion to an optical FFPC and its strongly enhanced emission on \nounderline{an} infrared (IR) transition into the cavity mode. Simultaneously, we measure the suppression of the spontaneous emission into free space on a strong ultra-violet (UV) transition of the same ion. While the presence of the cavity increases the ion's IR transition rate more than fourfold, the free space emission of the ion on the UV transition is suppressed by 66\%.
Employing spectroscopic methods and the measurements of optical pumping dynamics, all relevant experimental parameters are determined and used for successfully modelling the experimental results. 
\begin{figure}[t]
\begin{center}
\includegraphics[width=\columnwidth]{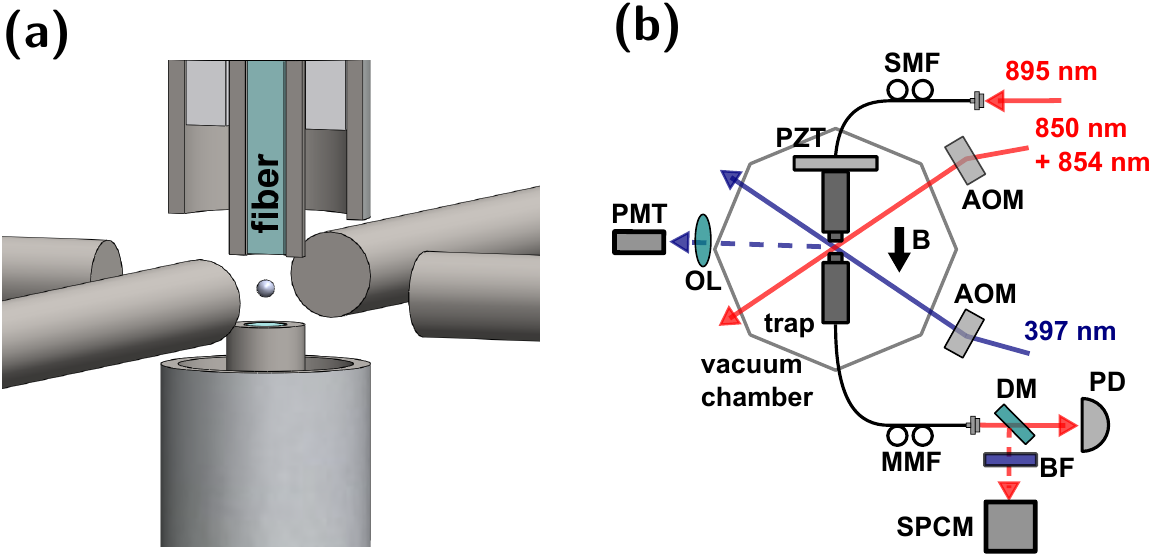}
\caption{\label{fig:fig_trap}(a) A close-up view of the ion trap structure with an integrated FFPC. Only a cross-section of the upper assembly is shown to reveal the internal structure: the fibers reside inside the inner electrodes and electrical isolation between the inner and outer electrodes is provided by ceramic spacer tubes. Additional four electrodes are located on the radial plane at the same height as the ion. (b) A simplified schematic of the experimental set up. (Key: \textbf{AOM}: acousto-optic modulator, \textbf{BF}: band-pass filter for 866 nm, \textbf{DM}: dichroic mirror, \textbf{MMF}: multi-mode fiber, \textbf{OL}: objective lens, \textbf{PD}: photo-detector for the transmission of the 895 nm beam, \textbf{PMT}: photo-multiplier tube, \textbf{PZT}: piezo-electric transducer, \textbf{SMF}: single-mode fiber, \textbf{SPCM}: single-photon counting module.)  The output of the MMF is filtered with a DM which transmits the 895 nm beam used for cavity-locking. The reflection is further filtered by a BF before being detected by the SPCM. The magnetic field (B) is controlled by external coils (not shown) and applied vertically along the cavity axis.
}
\end{center}
\end{figure}

A single $^{40}\mathrm{Ca}^+$ ion is trapped in an endcap style radio-frequency (rf) Paul trap described in \cite{Takahashi:13,Kassa:16}.  It is formed by a pair of electrode assemblies each consisting of two concentric stainless steel tubes (see \figref\ref{fig:fig_trap}(a)). For both assemblies, the outer electrode is recessed by 230 $\mu$m with respect to the inner electrode. The separation between the inner electrodes of the opposing assemblies is 350 $\mu$m. By applying an rf voltage at a frequency of 19.6 MHz to the outer electrodes while setting the inner \nounderline{electrodes} to rf-ground, a trapping potential is formed between the two electrode assemblies. The axial and radial secular frequencies are measured to be 3.46 MHz and 1.96 MHz, \nounderline{respectively}, with an estimated trap depth of 0.9 eV.
Four additional electrodes are placed in the radial plane at a distance of 1.0 mm from the center of the trap. By applying dc voltages to two of these radial electrodes as well as  the upper and lower inner electrodes of the main assemblies, stray electric fields are compensated to minimize excess micromotion of the ion by using the standard rf correlation technique \cite{Berkeland:98}.
The ion trap is combined with an FFPC by incorporating each fiber into the tubular inner electrodes. Both fibers have a CO$_2$ laser machined concave facet \cite{Takahashi:14} with radii of curvature of 560 $\mu$m and a high reflective coating of 25 ppm transmission at 866 nm. \nounderline{One is} a single mode fiber to serve as the cavity input and \nounderline{the other is} a multi-mode fiber which constitutes the output of the cavity. Each fiber is retracted by 5-10 $\mu$m from the end facet of \nounderline{the inner electrode it is inserted in}. The resulting cavity length is optically measured to be 367 $\mu$m which leads to an optimal coherent ion-cavity coupling of $g_0 = 2\pi \cdot 17.2$ MHz \nounderline{with} the $\mr{P}_{1/2} - \mr{D}_{3/2}$ transition of $^{40}\mathrm{Ca}^+$ at 866 nm. \nounderline{This is the theoretical expectation} \nounderline{when the ion-cavity overlap is optimal}. The cavity finesse is 48,000 corresponding to a cavity decay rate of $\kappa = 2\pi \cdot 4.2$ MHz. The magnetic field throughout the experiment is set to 0.78 G along the cavity axis.

\begin{figure}[t]
	\centering
	\includegraphics[width=0.8\linewidth]{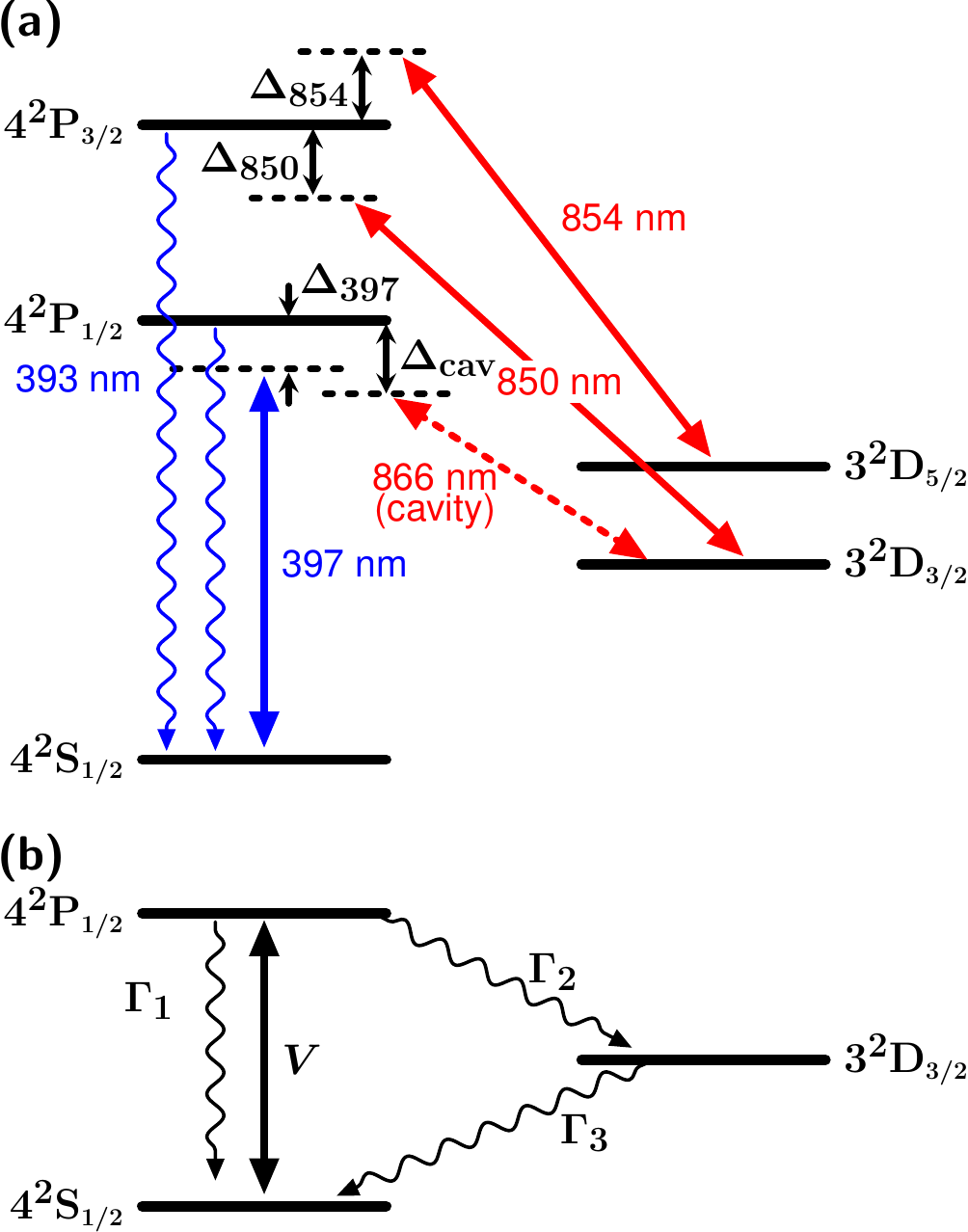}
\caption{(a) Energy level diagram of the \Ca ion and driving lasers with relevant detunings. The cavity field operates on the 866 nm transition between the $4^2\mr{P}_{1/2}$ and $3^2\mr{D}_{3/2}$ states. The wavy lines indicate the UV fluorescence emissions to be detected with the PMT. The detunings are measured positive\nounderline{ (negative)} when the laser is blue\nounderline{ (red)} detuned. (b) Effective scheme for the low-lying three levels.}\label{fig:energy-scheme}
\end{figure}

Fig.~\ref{fig:fig_trap}(b) shows a schematic of the experimental set up.
The ion is driven by three different laser beams at 397 nm, 850 nm and 854 nm which are near resonant with the $\mr{S}_{1/2}-\mr{P}_{1/2}$, $\mr{P}_{3/2}-\mr{D}_{3/2}$ and $\mr{P}_{3/2}-\mr{D}_{5/2}$ transitions respectively. Associated with these lasers are the detunings $\Delta_{397}$, $\Delta_{850}$ and $\Delta_{854}$ from the respective resonance frequencies as shown in \nounderline{Fig.}~\ref{fig:energy-scheme}. The laser at 397 nm is used for Doppler cooling the ion.
The fluorescence on this cooling transition as well as the spontaneous emission from the $\mr{P}_{3/2}$ state is detected with a photo-multiplier tube (PMT) via free space objective lenses. The repumper lasers at 850 nm and 854 nm are used to depopulate \nounderline{the} meta-stable $\mr{D}_{3/2}$ and $\mr{D}_{5/2}$ states and bring the ion back to the $\mr{S}_{1/2}$ state via spontaneous decay from the $\mr{P}_{3/2}$ state.
Another laser beam at 895 nm is injected into the FFPC through the input single-mode fiber and employed to stabilize the cavity length using the Pound-Drever Hall technique. The error signal for the cavity length is generated from the transmission signal through the multi-mode fiber and fed back to the piezo-electric transducer (PZT) attached to the upper assembly, which in turn changes the length of the FFPC. The frequency of this stabilizing laser is adjusted such that the FFPC satisfies a double-resonance condition for both the $\mr{P}_{1/2}-\mr{D}_{3/2}$ transition and the stabilizing beam. All the lasers are frequency-stabilized to a reference laser via a scanning cavity transfer lock \cite{seymour2010fast}.

\begin{figure}[t]
	\centering
	\includegraphics[width=0.6\linewidth]{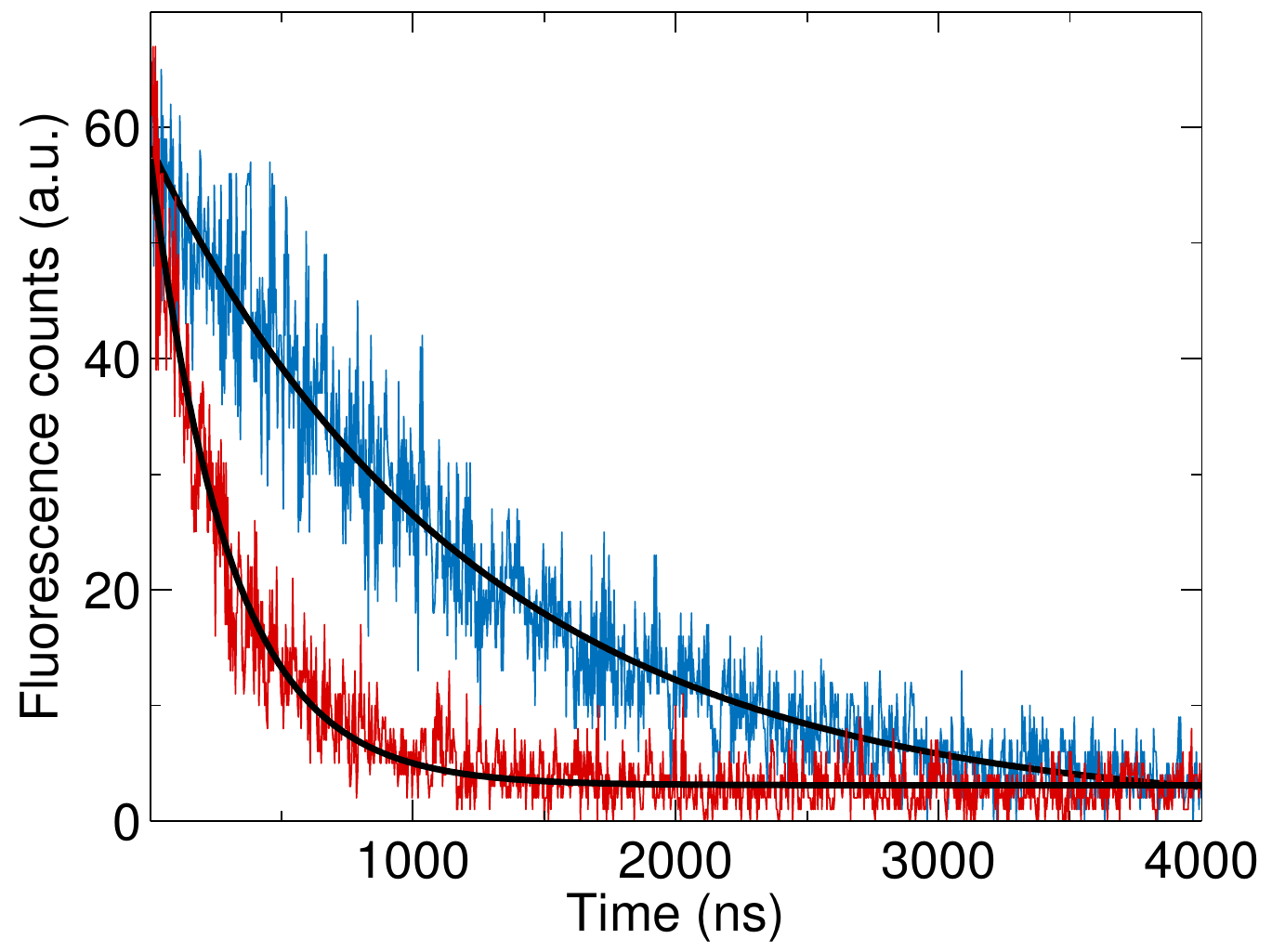}
	\caption{(a) UV Fluorescence detected with the PMT as the ion is shelved to the $\mr{D}_{3/2}$ state with (red) and without (blue) the cavity on the Raman resonance.  The black solid lines are exponential fits.}
	\label{fig:397shelving}
\end{figure}

A standard technique of observing the Purcell effect is through the measurement of the decay rate of the relevant transition. Here, it is manifested as an increased rate at which the ion's population is transferred from the $\mr{P}_{1/2}$ state to the $\mr{D}_{3/2}$ state. To infer this rate, we measure the transient change of the ion's fluorescence rate after the repumpers are abruptly switched off. \figref~\ref{fig:397shelving} shows the UV fluorescence in multiple repetitions of \nounderline{this} shelving \nounderline{process} with and without a near-resonant cavity. The detuning of the cavity, \nounderline{$\Delta_{\mr{cav}}$}, and \nounderline{that of the} cooling laser, \nounderline{$\Delta_{397}$}, are set to satisfy a Raman resonance condition $\Delta_{\mr{397}} = \nounderline{\Delta_{\mr{cav}}} = -2\pi\cdot 11.4$ MHz. Exponential decay fits to the data give time constants of $\tau_{\mr{on}} = 292 \pm 5$ ns and $\tau_{\mr{off}} = 1246 \pm 23$ ns for the on- and off-resonant cases respectively, demonstrating that the decay rate is enhanced by more than a factor of four by the cavity.

\begin{figure}[t]
	\centering
	\includegraphics[width=0.6\linewidth]{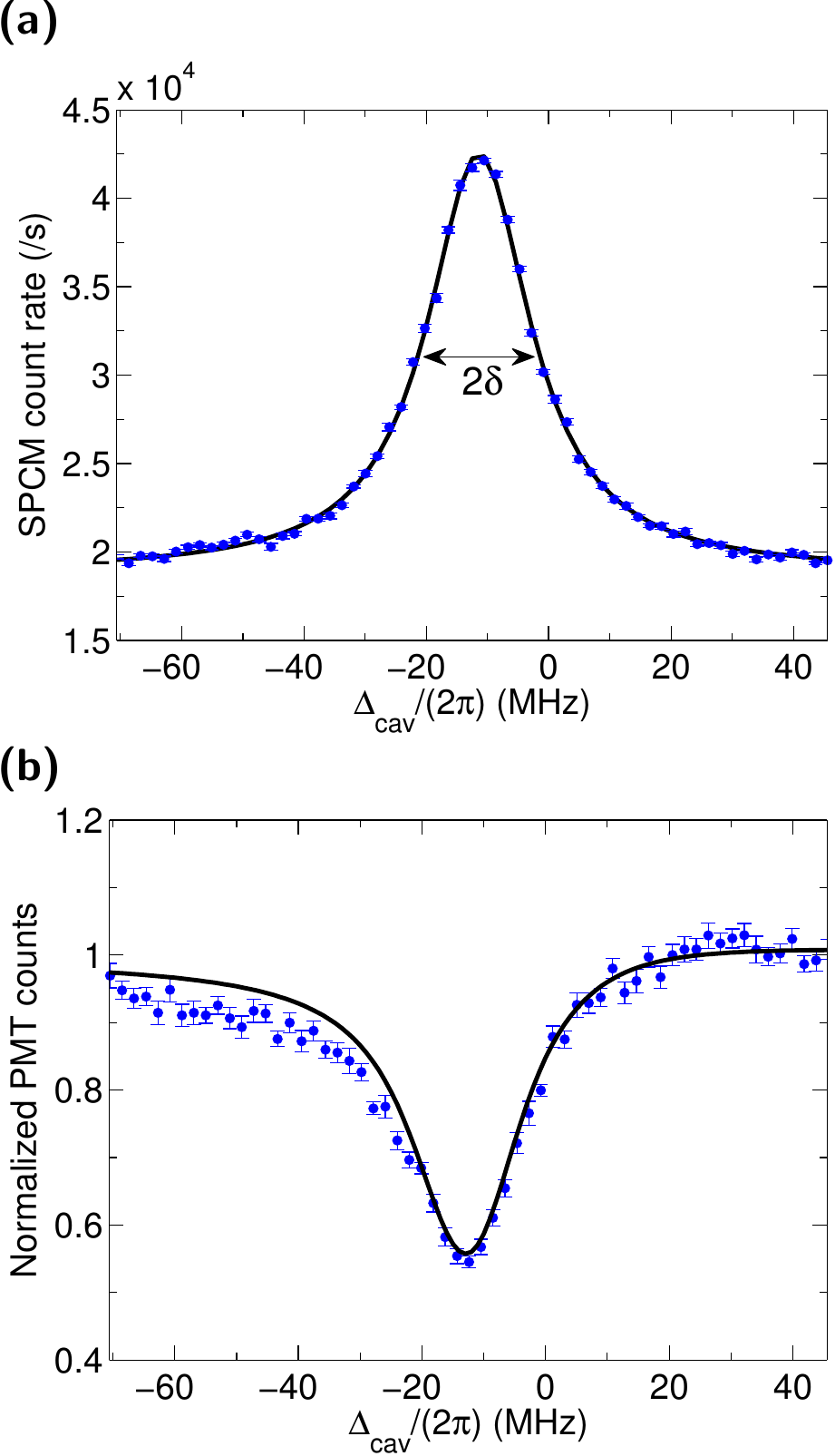}
	\caption{(a) Photon counting rate of the SPCM as the cavity detuning is scanned across the Raman resonance. Unfiltered leakage from the repumping  and locking lasers contribute to the background counts of $\sim$19,000 c/s. The black solid line is a fit by the Lorentzian function. The error bars show the statistical mean standard errors. (b) Normalized UV fluorescence measured simultaneously with (a).The black solid line is a numerical simulation using the measured experimental parameters.}
	\label{fig:scans-crop}
\end{figure}

However, the Purcell effect can also be observed in the interplay of decay rates between competitive transitions from the same excited state. 
While driving the ion continuously and scanning the cavity detuning across the Raman resonance, spectra for the cavity emission and the free-space UV fluorescence can be measured simultaneously with the SPCM and PMT respectively (see \figref\ref{fig:fig_trap}(b)). The cavity emission spectrum in \figref\ref{fig:scans-crop}(a) has a maximum measured net count rate of $\sim$22,000 c/s and a half width half maximum of $\delta = 10.3 \pm 0.1$ MHz. The UV fluorescence in \figref\ref{fig:scans-crop}(b) is normalized to the fluorescence rate measured with a far-detuned cavity. Anti-correlation between the two spectra is clearly visible.
The suppression of the UV fluorescence around the Raman resonance
can be understood by using rate equations for the effective model shown in the inset of \figref~\ref{fig:energy-scheme}. In this model only the populations in the low-lying three levels are considered and those in the $\mr{P}_{3/2}$ and $\mr{D}_{5/2}$ states are ignored. $\Gamma_1$ and $\Gamma_2$ are the spontaneous decay rates from the $\mr{P}_{1/2}$ state to the $\mr{S}_{1/2}$ and $\mr{D}_{3/2}$ states respectively, and $V$ is the pumping rate of the cooling laser. The 850 nm and 854 nm repumping lasers  are modelled as an effective incoherent decay from the $\mr{D}_{3/2}$ state to the $\mr{S}_{1/2}$ state characterized by decay rate $\Gamma_3$.
If the cavity modifies the decay rate $\Gamma_2$ to $\Gamma_2'$ due to the Purcell effect, the modified normalized fluorescence rate is given by 
\begin{align}
 \nounderline{\frac{N'_\mr{P}}{N_\mr{P}} \approx 1-\frac{1-v}{1+w\left(\frac{\Gamma_1+2V}{V}\right)}},
 \label{eq:norm-fluorescence-3level}
\end{align}
where $v = \Gamma_2/\Gamma_2'$, $w = \Gamma_3/\Gamma_2'$, and $N_\mr{P}$ and $N'_\mr{P}$ are the $\mr{P}_{1/2}$ state populations at equilibrium with and without the resonant cavity respectively. Here $\Gamma_1 + V \gg \Gamma_2$, which is approximately satisfied in our experiment, is used.
One can see that increasing $\Gamma_2'$ (decreasing $v$ and $w$) results in the suppression of the normalized fluorescence rate.

\begin{figure}[t]
 \centering
  \includegraphics[width=\linewidth]{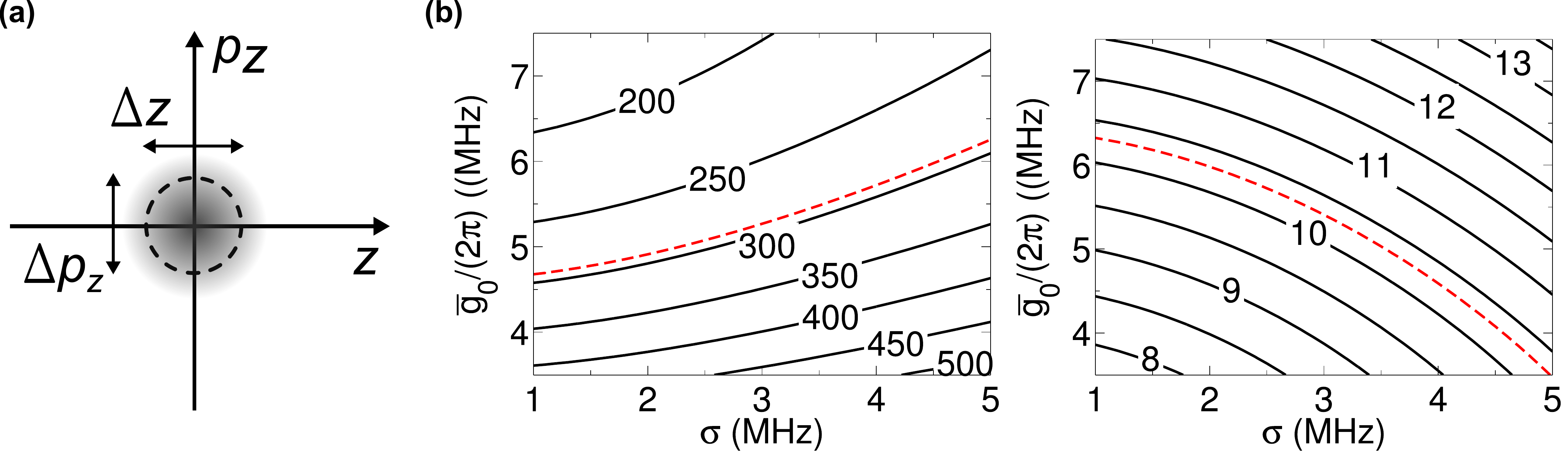}
  \caption{(a) The ion is in a thermal distribution in the phase space spanned by the spatial(=$z$) and momentum(=$p_z$) coordinates. The delocalization in $z$ results in the averaging of $g_0$ whereas the distribution along $p_z$ results in the spectral inhomogeneous broadening due to the Doppler effect. (b) Numerical simulation of $\tau_\mr{on}$ (left) and $\delta$ (right) as a function of $\bar{g}_0$ and $\sigma$ shown as 2D contour plots. The labels on the contour lines are in units of ns (left) and MHz (right). The red dashed contour lines correspond to the measured values of $\tau_\mr{on}$ = 292 ns and $\delta$ = 10.3 MHz}.
 \label{fig:phase_space+2Dsim}
\end{figure}

Even though qualitative understanding of the anti-correlated photon emissions can be obtained with the simplified model, in order to quantitatively analyze the measurement results, detailed characterization of the experimental parameters in conjunction with numerical simulation is necessary.
Our simulations are based on the master equation model involving all the relevant 18 Zeeman sublevels. The laser  parameters are measured individually and summarized in Table~\ref{tab:exp-params}. The detunings are obtained spectroscopically. 
Since no repumping dynamics is involved in the shelving measurement with the off-resonant cavity (\figref~\ref{fig:397shelving}), the only unknown parameter is $\Omega_{\mr{397}}$ for a given $\Delta_{\mr{397}}$. By fitting the numerical simulation to the measured time constant $\tau_{\mr{off}}$, the Rabi frequency $\Omega_{\mr{397}}$ is obtained.
Having deduced $\Omega_{\mr{397}}$, $\Omega_{850}$ is similarly obtained from a shelving measurement to the $\mr{D}_{5/2}$ state. $\Omega_{854}$ is obtained from the ac Stark shift caused by the 854 nm laser.
A key parameter to characterize our system is the ion-cavity coherent coupling strength $g_0$.
It can be extracted from the combination of the results in Fig.~\ref{fig:397shelving} and Fig.~\ref{fig:scans-crop}(a).
$g_0$ is a function of the ion's axial position ($=z$) in the standing wave of the cavity field ($\propto \cos(kz)$), and the finite spatial localization of the ion leads to the averaging of couplings at different positions \cite{begley2016optimized}.
Furthermore, the ion's motion as well as the instability of the cavity and laser locks introduce inhomogeneous broadening in the spectra shown in \figref\ref{fig:scans-crop}.
This can be described as a Gaussian distribution of the cavity detuning $\Delta_\mr{cav}$ with a standard deviation $\sigma$.
These two effects, the spatial average of $g_0$ and the inhomogeneous spectral broadening, can be described by the ion's spatial and momentum distributions in phase space respectively (see \figref\ref{fig:phase_space+2Dsim}(a)) and hence are taken into account separately.
In a simulation the former effect is taken into account by simply using an averaged effective coupling $\bar{g}_0$ rather than $g_0$.
The effect of inhomogeneous broadening is taken into account by calculating the weighted average of the simulations over a range of $\Delta_{\mr{cav}}$ with a width $\sigma$.
In order to deduce $\bar{g}_0$ and $\sigma$, we numerically calculate the dependence of $\tau_{\mr{on}}$ and $\delta$ on them, as shown in \figref~\ref{fig:phase_space+2Dsim}(b).
The dashed contour lines in \figref\ref{fig:phase_space+2Dsim}(b) correspond to the experimentally observed values of $\tau_{\mr{on}}$ and $\delta$ and the crossing point of these two lines uniquely determines the values of $\bar{g}_0$ and $\sigma$ in our experimental realization. As a result, we get $\bar{g}_0 = 2\pi\cdot (5.3 \pm 0.1)$ MHz and $\sigma = 3.1 \pm 0.2$ MHz. 

\begin{table}[t]
 \begin{tabular}{|c|c|c|c|}
  \hline
  $\Delta_{\mr{397}}$ & -$2\pi\cdot11.4 \pm 0.2$ MHz & $\Omega_{\mr{397}}$ & $2\pi\cdot18.2\pm 0.2$ MHz\\
  \hline
   $\Delta_{\mr{850}}$ & -$2\pi\cdot1.1\pm 0.1$ MHz & $\Omega_{\mr{850}}$ & $2\pi\cdot6.5\pm 0.1$ MHz \\
  \hline
  $\Delta_{\mr{854}}$ & $2\pi\cdot24.8\pm 0.1$ MHz & $\Omega_{\mr{854}}$ & $2\pi\cdot8.9\pm 0.4$ MHz  \\
  \hline
 \end{tabular}
 \caption{Measured laser parameters for the simulations in \figref\ref{fig:scans-crop}(c) and \figref\ref{fig:Rmin-vs-850}. $\Omega_{\mr{397}}$, $\Omega_{\mr{850}}$ and $\Omega_{\mr{854}}$ are the Rabi frequencies for the relevant laser beams. The cooling (397 nm) and repumping (850 nm and 854 nm) lasers are vertically and horizontally polarized respectively where the quantization axis is given by the vertically applied magnetic field.}
 \label{tab:exp-params}
\end{table}

\begin{figure}[t]
 \centering
  \includegraphics[width=0.65\linewidth]{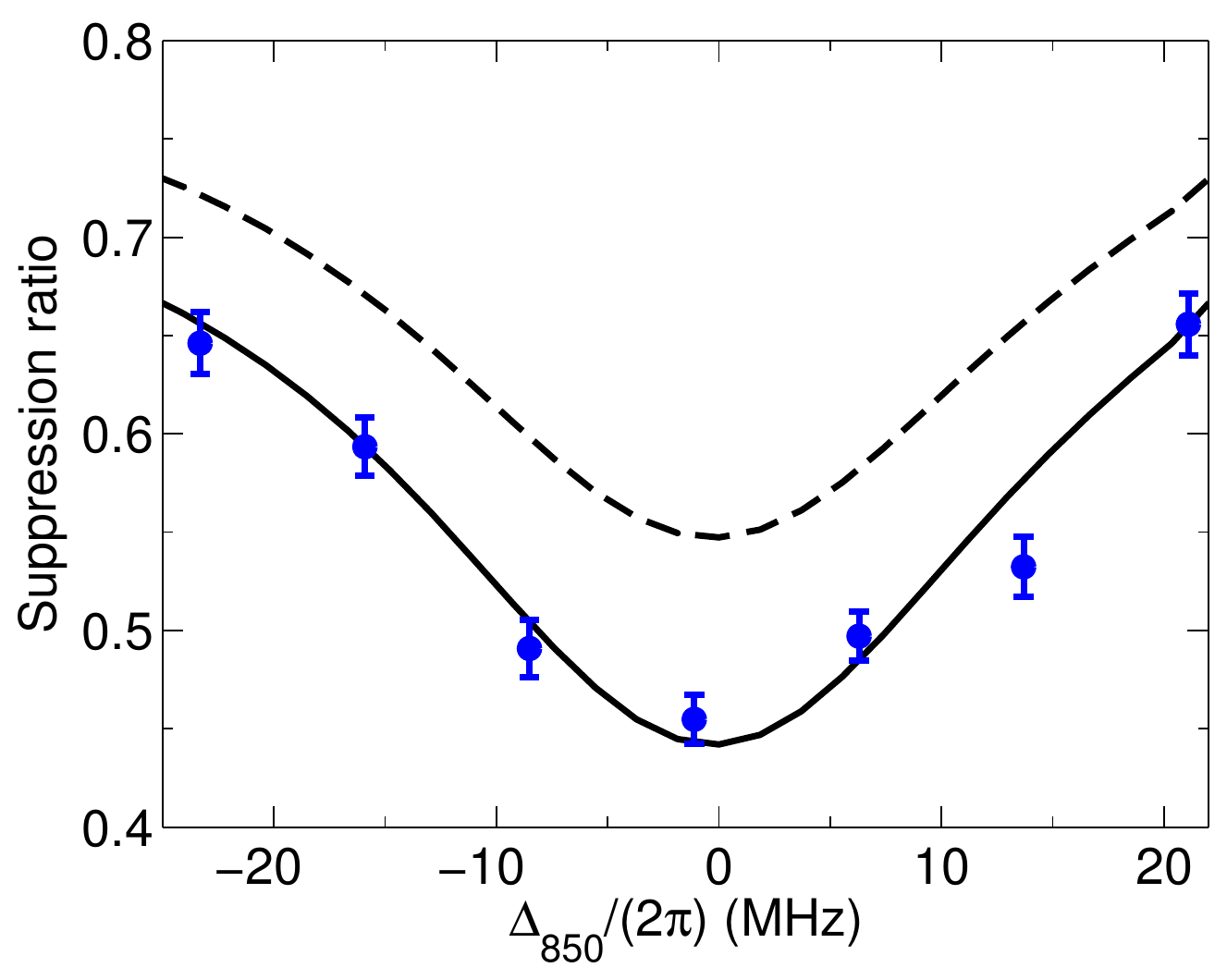}
  \caption{Maximum suppression in the normalized UV fluorescence spectra for various $\Delta_\mr{850}$. The black solid (dashed) line shows the numerical simulation including (excluding) the fluorescence at 393 nm. Apart from $\Delta_\mr{850}$, the same parameters as the ones in Table~\ref{tab:exp-params} are used.
 }
 \label{fig:Rmin-vs-850}
\end{figure}

Having obtained $\bar{g}_0$ and $\sigma$ in addition to the experimental parameters in Table~\ref{tab:exp-params}, numerical simulation for the UV fluorescence spectrum is carried out without a free fitting parameter (except for the frequency offset which is adjusted by using the peak position of the cavity emission). The theoretical prediction shown as the black line in \figref\ref{fig:scans-crop}(b) matches the experimental data well, demonstrating the validity of our model and the prior measurements. 
Further scans are taken with different 850 nm repumper detunings, $\Delta_{850}$, and the maximum suppression in the normalized UV fluorescence spectra are collected and shown in \figref~\ref{fig:Rmin-vs-850}. As the 850 nm laser is detuned, the effective repumping rate is decreased. This corresponds to decreasing $w$ in \eqnref(\ref{eq:norm-fluorescence-3level}), which leads to further suppression of the normalized fluorescence. As a result, a total suppression of the fluorescence up to 66 \% has been observed. This is a clear demonstration that the mere presence of a resonant cavity can significantly alter the radiation property of a single emitter. Note further that we detect the combined fluorescence at 397 and 393 nm (see \figref\ref{fig:energy-scheme}). If only the photons at 397 nm were detected, even greater suppression would be seen as shown in the dashed line in \figref\ref{fig:Rmin-vs-850}.  

In conclusion, we have developed an ion trap with an integrated high finesse FFPC and have successfully coupled a single ion to the cavity. Due to the $\Lambda$-type three-level structure in $^{40}\mathrm{Ca}^+$, the vacuum-stimulated emission of the resonant cavity on the $\mr{P}_\mr{1/2}$-$\mr{D}_\mr{3/2}$ transition leads to the strong suppression of the fluorescence on the $\mr{P}_\mr{1/2}$-$\mr{S}_\mr{1/2}$ transition. As a result, anti-correlated photon emission rates at two different wavelengths have been observed between the \nounderline{IR} cavity emission and the free-space UV fluorescence. From the thorough analysis of the measurement results, we have obtained an averaged ion-cavity coupling of $\bar{g}_0 = 2\pi\cdot (5.3 \pm 0.1)$ MHz with a corresponding cooperativity parameter of $C = \frac{\bar{g}_0^2}{2\kappa\gamma} = 0.30$.  This coupling strength is the largest reported value so far for a single trapped ion in the IR domain.
Currently the overlap between the ion and cavity field is limited by the construction accuracy with which the trap was built. 
However, by applying synchronous rf voltages on the radial electrodes, the rf null of the potential can be displaced to optimize the ion-cavity overlap without increasing excess micromotion \cite{herskind2009positioning}. This would greatly improve the ion-cavity coupling towards the strong coupling regime.

\begin{acknowledgments}
We gratefully acknowledge support from EPSRC through the UK Quantum Technology Hub: NQIT - Networked Quantum Information Technologies (EP/M013243/1 and
EP/J003670/1). We thank late Wolfgang Lange for his contribution in the foundations of this experiment. H.T. thanks Japan Science and Technology Agency (PRESTO) for their funding at the early stage of the experiment.
\end{acknowledgments}

\bibliographystyle{unsrt}
\bibliographystyle{apsrev4-1}
\bibliography{Purcell}

%
%
%
\end{document}